\documentstyle[a4]{article}

\begin{document}

\newcommand{\bib}{\bibitem}
\newcommand{\er}{\end{eqnarray}}
\newcommand{\br}{\begin{eqnarray}}
\newcommand{\be}{\begin{equation}}
\newcommand{\ee}{\end{equation}}
\newcommand{\epe}{\end{equation}}
\newcommand{\bea}{\begin{eqnarray}}
\newcommand{\eea}{\end{eqnarray}}
\newcommand{\ba}{\begin{eqnarray}}
\newcommand{\ea}{\end{eqnarray}}
\newcommand{\epa}{\end{eqnarray}}
\newcommand{\ar}{\rightarrow}
\newcommand{\dslash}{\partial\!\!\!/}
\newcommand{\aslash}{a\!\!\!/}
\newcommand{\pslash}{p\!\!\!/}
\newcommand{\bslash}{b\!\!\!/}
\newcommand{\kslash}{k\!\!\!/}
\newcommand{\rslash}{r\!\!\!/}
\newcommand{\cslash}{c\!\!\!/}
\newcommand{\fslash}{f\!\!\!/}
\newcommand{\Dslash}{D\!\!\!\!/}
\newcommand{\Aslash}{{\cal A}\!\!\!\!/}
\def\r{\rho}
\def\D{\Delta}
\def\R{I\!\!R}
\def\l{\lambda}
\def\D{\Delta}
\def\d{\delta}
\def\T{\tilde{T}}
\def\k{\kappa}
\def\t{\tau}
\def\f{\phi}
\def\p{\psi}
\def\z{\zeta}
\def\ep{\epsilon}
\def\hx{\widehat{\xi}}
\def\na{\nabla}

\date{}
\title{Bosonization and Duality in Arbitrary Dimensions: New Results. }
\author{M. Botta Cantcheff$^{a, b}$\thanks{{e-mail: botta@cbpf.br}} and J. A. Helayel-Neto$^{a, b}$
\thanks{{e-mail: helayel@gft.ucp.br}} \vspace{2mm} \\
{\small {\bf $^a$} Centro Brasileiro de Pesquisas Fisicas (CBPF-DCP),}\\
{\small Rua Dr. Xavier Sigaud, 150, Urca, 22290-180, RJ, Brasil.} \\
{\small {\bf $^b$} Grupo de F\'{\i}sica Te\'orica Jose Leite Lopes, Petr\'opolis, RJ, Brazil.}}

\maketitle

 \begin{abstract}

 A generic massive Thirring
 Model in three space-time dimensions exhibits a correspondence
 with a topologically massive bosonized gauge action associated to a self-duality constraint, and we write down a general expression for this relationship.

 We also generalize this structure to $d$ dimensions, by adopting
   the so-called doublet approach, recently introduced.
      In particular, a non-conventional formulation of the bosonization technique in higher
       dimensions (in the spirit of $d=3$), is proposed and, as an
        application, we show how fermionic (Thirring-like) representations for
 bosonic topologically massive models in four dimensions
%(such as Cremmer-Scherk-Kalb-Ramond and Born-Infeld-Kalb-Ramond)
 may be built up.

\end{abstract}

 \section{Introduction}

This paper has a two-fold purpose: to establish both bosonic
first-order (gauge non-invariant)
  and fermionic Thirring-like formulations for very general topologically massive theories
\cite{Cremer,Kalb,bf} in arbitrary dimensions. We show these correspondences by extending the
    techniques typically used for duality and bosonization in three-dimensional models via
     the doublet-formalism \cite{dob,tmdob}, which
      appear insensitive to the  space-time dimensionality.

Duality has a fundamental importance in our understanding
of various non-perturbative aspects of point-particle and string theories.

Some years ago \cite{DJ}, Deser and Jackiw
developed the concept of parent action approach
 \cite{suecos} and showed duality
 between the so-called self-dual
   theory (SD)\cite{TPvN} in three dimensions
 and the topologically massive gauge theory,
     referred to as Maxwell-Chern-Simons (MCS). Furthermore, it was shown
     \cite{chap1}
     that the SD model is connected, via the so-called bosonization
     technique, to the Thirring model, \be S^{(ferm)}(\psi,{\bar \psi}) \equiv \int d^3x \left(
\bar\psi(i\dslash - m) \psi -\frac{g^2}{2} j^{\mu}j_{\mu} \right),
~~~~~j^{\mu}\equiv \bar\psi \gamma^\mu \psi . \ee Bosonization is
the mapping of a quantum field theory for interacting fermions
onto an equivalent theory for interacting bosons \cite{bos}.

 Recently, Tripathy and Khare \cite{thrip} considered a modification of this model
 by replacing the Maxwell term by $\sqrt{1-F^2}$,
the Born-Infeld Lagrangian \cite{infeld}. Bosonization and
dual-correspondences of its topologically massive version, the
Born-Infeld-Chern-Simons theory \cite{gibbons,I1}, have recently been
 studied motivated by the fact that these theories naturally
    appear in the context of Dp-branes \cite{born} whose dynamics
 is described by Born-Infeld-Chern-Simmons-actions in $d=(p + 1)$ dimensions. In
        particular, the D2-brane is described by the
         3d-Born-Infeld-Chern-Simmons model. This result is one of the main motivatons for the study carried out in our paper,
whose purpose is precisely the extension of the above result to a general dimension $d$.
Despite the notion of self-duality in arbitrary dimensions introduced
in Ref. \cite{dob}, which has proven to be a crucial hint in order to establish dual correspondences \cite{tmdob},
 this extension is not set in a straightforward way. This becomes clearer mainly in Section 4,
where non-conventional fermionic currents must be introduced in
order to describe topologically massive models as purely fermionic
theories. Indeed, we succeed in setting up fermionic
representations for the topologically massive
Cremmer-Scherk-Kalb-Ramond model in four space-time dimensions and
also for more general gauge models, for instance involving a
Born-Infeld theory topologically coupled to a Kalb-Ramond field
(This theory shall be referred to as Born-Infeld-Kalb-Ramond).
Some interesting technical particularities also appear when the
bosonization procedure, initially thought for $d=3$ \cite{chap1},
is reproduced for $d=4$.

The main goal of this paper is thus to focus all these issues in a more general context.

We shall to construct this generalized framework by investigating two principal
types of extension for this structure: to consider arbitrary ($d$)
                dimensions and more general non-linearities
                 (arbitrary functions of the squared field-strength).

   This work is organized as explained
below. In Section 2, we briefly review the bosonization of the
Thirring model in three dimensions into a SD-model and the SD-MCS
duality. In Section 3, we generalize this to arbitrary non-linearities in the
Maxwell term: we show that this is always equivalent to a SD-model
in a generalized sense and find to a formula to relate the
theories of this correspondence.
Afterwards, we use a direct procedure to bosonize a generic Thirring model
with an arbitrary current-current coupling and connect it to the non-linearity of its
bosonic representations.

 Generalization of this structure to higher dimensions
 is the
matter of Section 4. We shall show in this section (for the particular case $d=4$, but
indicating the way for generalizing to higher dimensions
elsewhere), that bosonization may be implemented {\it in the same
way} as in 3d, via the recently introduced doublet
formalism\cite{dob,tmdob}, resulting in an alternative formulation of the
bosonization technique in four dimensions \cite{banerj}. Such as in the $3$-d case,
fermionic models bosonize to topologically massive ones; in particular, we
concentrate our discussion in specially interesting topologically massive
gauge theories in four dimensions: Born-Infeld-Kalb-Ramond and Cremmer-Scherk-Kalb-Ramond
 \cite{Cremer,Kalb,bf}.

 Finally, in Section 5, we draw
our general conclusions and emphasize on the aspects that concern
generalization to arbitrary dimensions.

%%%%%%%%%%%%%%%%%%%%%%%%%%%%%%%%%%%%%%%%%%%%%%%%%%%%%%%%%%%%%%%%%%%%%%%%%%%%%%%%%%%%%%%%%%%%%%

\section{A Short Introductory Review.}

Let us briefly review how the low-energy sector of a theory of
massive, electrically-charged, self-interacting fermions (the
massive Thirring Model) in (2 + 1)-dimensions may be bosonized
into a gauge theory, the Maxwell-Chern-Simons gauge theory
\cite{DJ,chap1}.

{\bf SD-MCS duality:}

In 2+1-dimensions, one currently defines the Hodge-Duality
operation by, \be \label{1} \mbox{}^{\star} a_\mu
=\frac{1}{m}\, \epsilon_{\mu\nu\lambda}\,\partial^\nu
a^\lambda\,, \ee
 where $m$ is a parameter that renders the
 $\mbox{}^{\star}$-operation
 dimensionless.

We refer to {\it self(anti-self)-duality} whenever the relations
$\mbox{}^{\star} a = + a \, , \, -a $, are respectively satisfied.
Throughout this paper, we shall introduce a parameter $\chi=\pm 1$
to express this self/anti-self-duality.

The so-called Self-Dual Model (Townsend, Pilch and van
Nieuwenhuizen \cite{TPvN}) is described by the following action,

\be
\label{180}
 S_{SD}(a)= \int\, d^3x\:\Bigg(\frac {\chi }{2m}\,
\epsilon_{\mu\nu\lambda}\, a^\mu\,\partial^\nu a^\lambda -
\frac{1}{2}\, a_\mu a^\mu \Bigg)\,.
\ee

The equation of motion is the self-duality relation:
\be
\label{190}
a_\mu =\frac{\chi}{m}\,
\epsilon_{\mu\nu\lambda}\,\partial^\nu a^\lambda\,.
\ee
This model is claimed to be chiral, and the chirality results
 defined precisely by this self-duality.

On the other hand, the gauge-invariant combination of a
Chern-Simons and a Maxwell term: \be \label{mcs}
S_{MCS}[A]= \int\, d^3x\:\Bigg(\frac{1}{4m^{2}}
F^{\mu\nu}F_{\mu\nu} -
\frac{\chi}{2m}\,\epsilon^{\mu\nu\lambda}\,A_{\mu}\,
\partial_{\nu}A_{\lambda}\Bigg)\,,
\ee
is the topologically massive theory, which is known to be equivalent \cite{DJ} to
 the self-dual model (\ref{180}).
$F_{\mu\nu}$ is the usual Maxwell field strength,

\be \label{285} F_{\mu\nu}[A] \equiv \partial_{\mu}A_{\nu} -
\partial_{\nu}A_{\mu}\, = 2\partial_{[\mu}A_{\nu]}. \ee

This equivalence may be verified with the parent action approach \cite{suecos}.
We write down the general
 parent action proposed by Deser and Jackiw in \cite{DJ}, which proves this equivalence:

 \be
 \label{masterDJ0}
 {\cal S}_{Parent}[A, a]= \chi\, {\cal S}_{CS}[A] -\int\,d^{3}x\,
  \left[ \epsilon^{\mu\nu\lambda}\,
F_{\nu\lambda} [A]
 a_{\mu} \, +  m \, a_{\mu } a^{\mu } \right] ,
\ee

where
\be
\label{CS}
{\cal S}_{CS}[A] \equiv \int\,d^{3}x\, \epsilon^{\mu\nu\lambda}\, \left(
{A}_\mu\partial^{}_{\nu}A_{\lambda} \right),
\ee
is the Chern-Simons action \cite{Deser}\footnote{In fact, vaying this action
 with respect to $f$,
and eliminating this in the action from the equation of motion,
one get
 $S_{SD}(a^{\mu})$.
Varying ${\cal S}_{Parent}$ with respect to $a$, we obtain

\be
 a^{\mu} = -{1 \over 2 m }\epsilon^{\mu\nu\lambda}\,
F_{\nu\lambda}[A];
 \ee
plugging this back into (\ref{masterDJ0}), and using \be
\epsilon^{\mu\nu\alpha}\epsilon_{\mu\nu\lambda} =
2\,\delta_\lambda^\alpha \label{ep} ,\ee we recover the
MCS-action, Eq. (\ref{mcs}).}.

{\bf Bosonization and Thirring-MCS correspondence:}

On the other hand, the (Euclidean) fermionic partition function
for the three-dimensional massive Thirring reads as below: \be
\label{BB101}
 Z^{(ferm) } = \int \,{\cal D}\bar\psi {\cal D}\psi\; e^{ -\int \left( \bar\psi(\dslash
+ m) \psi -\frac{g^2}{2} j^{\mu}j_{\mu}
\right ) d^3x } ,
\ee
with the coupling constant $g^2$ having dimensions of inverse mass
and $m$ is the fermion mass.

It is well-known that this model can be bosonized to the self-dual
model \cite{chap1},  \be \label{zthsd} Z^{(ferm)} \approx Z^{SD},
\ee in the
 low-energy limit.

Thus, thanks to the equivalence between (\ref{180})  and
(\ref{mcs}), one can establish the following bosonization identity:

\be
Z^{(ferm) } \approx Z^{MCS}.
\ee
This equation, together with (\ref{zthsd})
, both connected by SD-MCS duality (\ref{masterDJ0})
, constitutes the kernel of this work:
our main purpose is to actually study the generalizations
 of this structure along two {\it independent}
lines:

\begin{itemize}

\item for Thirring-like models with an arbitrary current-current
coupling: correspondence rule with self-dual and non-linear topologically
 massive theories.

\item for arbitrary dimensions:
  fermionic Thirring-like models in general dimensions correspond to
   topologically massive theories, such as in 3d
\footnote{In general dimensions, the Abelian gauge field
generalizes to a pair of field forms.}.

\end{itemize}

Clearly, both generalizations are suitable to be connected to one another.

%%%%%%%%%%%%%%%%%%%%%%%%%%%%%%%%%%%%%%%%%%%%%%%%%%%%%%%%%%%%%%%%%%%%%%%%%%

 \section{Duality between Non-Linear
  Self-Dual and Topologically Massive Models in Three Dimensions.}

In this section, we shall generalize the correspondence SD-MCS to
account for arbitrary non-linearities. We will show here that the
TM model with non-linearity described by a function
$U(F^2)$\footnote{When $U$ is linear the theory is commonly
referred to as MCS.},

\be \label{370} S_{U(F^2)}[A]= \int\, d^3x \: \Bigg( U(F^{\mu\nu}F_{\mu\nu}) -
{\chi}\,\epsilon^{\mu\nu\lambda}\,A_{\mu}\,
\partial_{\nu}A_{\lambda}\Bigg)\, ,
\ee
corresponds to the also general Non-Linear Self-Dual model, with
non-linearity given by a potential $V(a^2)$:

\be \label{180-V1} {\cal S} _{V(a^2)}[a] = \int d^3
x\: V \left(a_\mu a^\mu \right) - \chi {\cal S}_{CS}[a]\, ,
\ee which is the non-linear version of the self-dual action
introduced
 in \cite{TPvN}. We shall refer to this theory as Non-Linear Self-Dual
Model.

It is useful to briefly clarify why the property of self-duality can
be attributed to this model. The equations of motion derived from
Eq.(\ref{180-V}) are given by
    \begin{equation}
    a_{\mu} = \frac{\chi}{2\,V^{\prime}}\,
    \epsilon_{\mu\nu\lambda}\,\partial^{\nu}a^{\lambda}\,,
    \label{AIS01B}
    \end{equation}
    where the prime denotes a derivative with respect to the argument.
    This non-linear SD model possesses a well-defined self-dual
    property in the same manner as its linear counterpart. This can be seen as
    follows. Define a field, ${}^{\star}a_{\mu} $, dual to $a_{\mu}$ as
    \begin{equation}
    \label{AIS01E}
    {}^{\star}a_{\mu} \equiv \frac{1}{2\,V^{\prime}}\,\epsilon_{\mu\nu\lambda}\,
    \partial^{\nu}a^{\lambda}\, ,
    \end{equation}
    and repeat this dual operation to find that, as
     consequence of the equations of motion (\ref{AIS01B}),
    \begin{equation}
    \label{AIS01F}
    {}^{\star}\left({}^{\star}a_{\mu}\right) = a_{\mu}.
    \end{equation}
Dual correspondences for
 this type of non-linear systems
 have recently been studied in the particular case of Born-Infeld
\cite{I1} and also in other specific cases in Ref. \cite{w}
(for instance, a power law $U(z)= z^r \, , \, r \epsilon Q $); which
use a method recently proposed \cite{w0} based on the
traditional idea of a local lifting of a global symmetry that may
be realized by iterative embedding Noether counter-terms.

 These approaches treat the non-linearities by introducing auxiliary
fields. In this section, we are going to confirm
the previous results by adopting the parent action approach and generalize them
further {\it without} introducing auxiliary fields; of course,
this enforces the evidence in favour of this so-called gauging
Noether method \cite{w0} as a useful dualization procedure.

To derive our results, we consider the following non-linear
generalization of the Deser-Jackiw Parent Action \cite{DJ}: \be
 \label{masterDJ}
 {\cal S}_{Parent}[A,a]= \chi{\cal S}_{CS}[A] -\int\,d^{3}x\,
  \left[ \epsilon^{\mu\nu\lambda}\,
F_{\nu\lambda}^{} [A]
 a_{\mu}^{} \, +  V (a_{\mu}^{} a^{\mu})\right] .
\ee

Varying it with respect to $A$,

\be
\label{201} \epsilon_{\mu\nu\lambda}\,
                  \partial^\nu [A^\lambda - a^\lambda ]=0 \,,
\ee

we write its solution as \be A^\lambda = a^\lambda +
\Delta^\lambda ,\ee
where $\Delta^\lambda =\partial^\lambda \Delta$ is pure gauge.
 Putting this back into (\ref{masterDJ}), we
recover $S_{V(a^2)}[a]$, equation (\ref{180-V1}).

Now, strictly following the standard program of the master action
approach \cite{suecos}, we must vary the parent action with
respect to $A$, and use the resulting equation to solve $A$ in
terms of the other field, $a$. Finally, one shall eliminate $A$
from the parent action.

Varying ${\cal S}_{Parent}$ with respect to $a$, we obtain
 \be
 \label{solf}
- 2 V'(a^2) a^{\mu }=\epsilon^{\mu\nu\lambda}\,
F_{\nu\lambda}^{}[A],
 \ee
from which it follows that
\be
 \label{solu1}
-2 a^2 V'(a^2)= a_\mu
\epsilon^{\mu\nu\lambda}\,F_{\nu\lambda}^{}[A] \ee

and
 \be \label{solu2}
\epsilon^{\mu\nu\lambda}\,
F_{\nu\lambda}^{}[A]\;\epsilon_{\mu\rho\alpha}\,
F_{\rho\alpha}^{}[A]= 2 F^2 = 4 a^2  [V'(a^2 )]^2 ;
 \ee
Formally, one can solve this for $a^2$ in terms of $F^2[A]$, and
put the result back into (\ref{masterDJ}) to express this action
in terms of the field $A$, which results to be a TM-theory .
Defining a function $W$ trough its inverse (whenever it exists),
\be W^{-1}(v) \equiv 2 v \, (V'(v))^2 ,~~~~~~v\ep\R, \ee
and substituting in the parent action by eq. (\ref{solu1}),
 we recover the generalized non-linear topologically
massive theory; the gauge invariant combination of a Chern-Simons
with a non-linear Maxwell term

 \be \label{3700} S_{MCS}[A]= \int\,
d^3x\:\Bigg( U(F^{\mu\nu}F_{\mu\nu}) -
\chi\,\epsilon^{\mu\nu\lambda}\,A_{\mu}\,
\partial_{\nu}A_{\lambda}\Bigg)\,,
\ee

where the functional $U$ is related to $V$ (that characterizes the
non-linearity of the self-dual model) by the formula:\be \label{UV}
U(q)=-2 W(q) V'(W(q)) + V(W(q) ), \ee with $q \ep \R^+$.

At the end of the next section, we shall mention some more
relevant examples of solutions to this equation.

%%%%%%%%%%%%%%%%%%%%%%%%%%%%%%%%%%%%%%%%%%%%%%%%%%%%%%%%%%%%%%%%%%%%%%%%%%%%%%%%%%%%%%%%%%%%%%%%
\subsection{ Bosonisation of Thirring models with
 arbitrary (current-current) coupling in $d=3$.}

In this section, we are going to find bosonization identities
 for the most general Thirring (fermionic) model,
 i.e. with an arbitrary
dependence on the current-current coupling; this remarkably
corresponds to a version of the MCS with {\it the same} dependence
on the square of the field strength \cite{sor}. We use a direct
procedure such as in the traditional case (Eq. (\ref{BB101})).

The particular
 case of Born-Infeld-Chern-Simons has already been studied in
 \cite{thrip,I1};
  clearly, these results are contained in the scheme presented here.

In fact, consider a generalization of the Thirring model to have a
term depending arbitrarily on
 $j^{\mu}$.
By relativistic invariance, the only possibility is the
generalized non-linear model:

\be
\label{BB10}
 Z_{T(j^2)}^{(ferm)} = \int \,{\cal D}\bar\psi
  {\cal D}\psi\; e^{-\int \left(8\pi \,\bar\psi(\dslash
+ m) \psi \,- \,T( {j^{\mu}j_{\mu}\over 2} ) \right) d^3x} , \ee
where the function $T$ is analytic and real-valued.

Next, we eliminate the non-linear interaction by introducing a
vector field, $a^{\mu}$, and using the identity: \be
e^{\int\!d^3\!x\, T( \frac{j^{\mu}j_{\mu}}{2})} =
\int{\cal D} a_{\mu} e^{-\int\!d^3\!x\, tr(V(
a^{\mu}a_{\mu}) + j^{\mu}a_{\mu})}, \label{NB20} \ee
where $V$ is related to $T$. We shall find this relation
 varying the exponent of the RHS with respect to $a$
to obtain:
 \be
 \label{solft}
- 2 V'( a^{\nu}a_{\nu})  a^{\mu } = j^{\mu}   ;
 \ee
from which there follow the relations,
\be
 \label{solu1t}
-2 a^2 V'(a^2)= a_\mu j^{\mu},
\ee
and
\be
\label{solu2t}  j^{\mu}j_{\mu}= 4 a^2 V'(a^2) .
 \ee

In principle, one can solve for $a$ (or $a^2$) from (\ref{solu2t})
in terms of $j^2$, and put the result back into (\ref{NB20}) to
express this action in terms of the current $j$, and recover the
non-linear Thirring model . Let us define again the function $W$ trough its
inverse, assuming it to be: \be W^{-1}(v)= 2 v
[V'(v)]^2 ;\label{34}\ee therefore, $W(q) =v$. Plugging these
equations back into (\ref{NB20}),
 we recover the generalized
non-linear Thirring Model, eq. (\ref{BB10}), where $T$ is given by
\be\label{35} T(q)= - 2 W(q)  V'(W(q) ) \, +  V(W(q) ) \, , \ee

Notice that, by virtue of (\ref{34}), eq. (\ref{35}) coincides
with (\ref{UV}); then, one obtains:

\be \label{TU} T(q)=U(q) ,~~~~~~~~~~q\ep\R^+,\ee

in agreement with the formal correspondence rule \cite{sor}, $j
\to \mbox{}^{*} F$ \footnote{$()\mbox{}^{*}$ is the usual Hodge's
operation}. This shall provide us with a general correspondence
bosonization identity for very general Thirring-like models and
topologically massive gauge theories.

 So, thanks to these results, one can represent the Thirring model as:

\be {\cal Z}_{T(j^2 /2)}  = \int {\cal{D}} a_{\mu}
\; \det (i\dslash + m + \aslash) \, e^{- \!\int d^3
x\: V \left(a^{\mu}a_{\mu}\right)}  , \label{NB30} \ee

Now, we proceed in the same way as in the typical case ($T(j^2/2)$,
linear) to evaluate the determinant. The determinant of the Dirac
operator is an unbounded operator and requires regularization.

 For $d=3$, the actual computation of this determinant will give
parity-breaking and
 parity-preserving terms that are computed in powers of the inverse mass,
\be
\ln \det (i\dslash + m + \aslash) =   \frac{\chi}{16\pi} {\cal
S}_{CS}[a] +  I_{PC}[a] + O(\partial^2/m^2)  \: . \label{NB40} \ee
Here, ${\cal S}_{CS}$ is given by

\be \label{NB50} {\cal S}_{CS}[a] = \int d^3 x\: i
\epsilon^{\mu\nu\lambda}\, (F_{\mu \nu} a_{\lambda} ) ;\ee it is
the Abelian Chern-Simons action, and the parity-preserving
contributions, to first-order, lead to the Maxwell action

\be
I_{PC}[a] = - \frac{1}{24\pi  m}\, tr\int d^3 x\: F^{\mu\nu} F_{\mu\nu}  .
\label{NB60}
\ee

In the low-energy regime, only the Chern-Simons action survives
yielding a closed expression
 for the determinant:

\be \ln \det (i\dslash + m + \aslash) =   \frac{\chi}{16\pi} {\cal
S}_{CS}[a] + o(m^{-1}) \label{NB75} \ee Using this result, we may
write:

\be \lim_{m\to\infty} {\cal Z}^{(ferm)}_{T(j^2)/2} =
  \int {\cal{D}}
a_{\mu}  \exp(-{\cal S}_{V(a^2)} [a])  , \label{NB80} \ee where
${\cal S}_{V(a^2)} $ is the non-linear version of the self-dual
action  introduced
 in \cite{TPvN},

\be \label{180-V} {\cal S}_{V(a^2)} [a] = \int d^3
x\: V \left(a_\mu a^\mu \right) -
  \chi {\cal S}_{CS}[a]
\ee Therefore, to the leading order in $1/m$, we have established
the identification with the Non-Linear Self-Dual theory:
 \be{\cal Z}^{(ferm)}_{T(j^2)/2} \approx {\cal Z}^{}_{V(a^2)} . \ee

 Finally, recalling that the model with dynamics defined by the
non-linear self-dual action (${\cal S}_{V(a^2)}$) is equivalent to
a non-linear Maxwell-Chern-Simons theory (${\cal S}_{U(F[a]^2)}$),
we use the relation (\ref{TU}) to establish the Bosonization
identity of the non-linear massive Thirring model with the
topologically massive theory along with the remarkable
identification of the potentials, as

\be {\cal Z}^{(ferm)}_{U(j^2/2)} \approx  {\cal Z}^{}_{U(F^2)} \:
. \label{NB100} \ee In some cases, it is relatively simple to
solve the equation (\ref{34}) (or, by virtue of (\ref{TU}), eq.
(\ref{UV})). Let us illustrate this by mentioning some relevant
examples: Taking a Thirring model with current-current interaction
described by a function $T(j^2) \propto ( j^{\mu}j_{\mu})^k$,
then, it is equivalent to a self-dual model with non-linearity
described by another power law: $V(a^2)\propto
(a^{\mu}a_{\mu})^{{k\over(2k-1)}}$, and by virtue of (\ref{TU}), the
corresponding model has a Maxwell term substituted by $U(F^2) \propto (
F^{\mu\nu}F_{\mu \nu})^k$. A simple inspection shows that this
result agrees with the one obtained in \cite{w}, which enforces the
validity of the method proposed there.

The Born-Infeld-Chern-Simons example sets a special case
since, as it can be directly verified from eq. (\ref{UV}), the
functional forms of the three models, {\it coincide}
\cite{thrip,w,w0} ; i.e $ T(q) = U(q) \propto V(q) \propto
\sqrt{1-(const\cdot q^2)}$, for all $q\ep\R$.

%%%%%%%%%%%%%%%%%%%%%%%%%%%%%%%%%%%%%%%%%%%%%%%%%%%%%%%%%%%%%%%%%%%%%%%%%%%%%

\section{General dimensions: Born-Infeld-Kalb-Ramond and Cremmer-Scherk-Kalb-Ramond
gauge theories.}

In this section, by considering doublets of field-forms, we show how the
structure described above may also be established in $d$
dimensions.

For general dimensions, it is possible to define self (and
anti-self)-duality for pairs (doublets) of form-fields with
different ranks \cite{tmdob}, close in spirit to the self-duality in
$(2+1)$-dimensions due to Townsend, Pilch and van Nieuwenhuizen
\cite{TPvN}. Remarkably, as it has been shown in ref \cite{tmdob},
the actions which describe this doublet-self-duality, result to be
{\it dual-equivalent} to topologically massive theories in $d$
dimensions, which involve BF-terms (topological coupling between
different Abelian gauge forms \cite{Cremer,bf} ); in the same way
that the SD-MCS duality (in $(2+1)$).

In this work, this parallel shall be enforced and generalized with
novel consequences on bosonization in high dimensions; besides,
new dualities between theories shall be established.

\vspace{0.5cm}

Let us consider a $d$-dimensional space-time with signature $s$ \footnote{i.e,
this is the number of minuses occurring in the metric}: we consider the tensor doublet,
\be
{\cal F} :=(f_{\mu_1 \cdots\mu_p} ,  g_{\mu_1 \cdots\mu_{d-p-1}} ), \ee
 where $f$ is
a $p(<d)$-form ( a totally antisymmetric tensor type $(0;p)$ ) , and $g$ is a $(d-p-1)$-form.
 ${\cal F} $ is an element of
the space $\Delta_p \equiv \Lambda_p \times \Lambda_{d-[p+1]}$
\footnote{A more detailed discussion on this construction and its motivations may be found in
 Refs. \cite{dob,tmdob}; however, the notion presented here is sufficient to make this paper self-contained. }.

There is a well defined notion of self (and anti-self)-duality for
the objects in this space based on the standard Hodge´s operation,
$()\mbox{}^{*}$ \footnote{For a  generic $q$-form,  $A$, the Hodge
dual is defined by \be (\mbox{}^{*} A )^{\mu_{q+1}\cdots \mu_{d}}
= \frac 1{q!} \; \epsilon^{\mu_1\cdots \mu_{d}} A_{\mu_{1}\cdots
\mu_{q}}.\ee} in a fashion extremely similar to the
$(2+1)$-dimensional case described above.  Consider the action
with topological coupling: \be \label{DSD} S_{DSD}[{\cal F}]
\equiv \int dx^d \,\left[ {-1\over m} g_{\mu_1\cdots\mu_{d-p-1}}
\epsilon^{\mu_1\cdots \mu_{d}} \partial_{\mu_{d-p}}
f_{\mu_{d-p+1}\cdots\mu_{d}} + \rho({\cal F}) \right] ,\ee where
$\rho({\cal F}) $ collects the explicit mass terms as, \be
 \rho({\cal F}) \equiv \frac 12 ([p+1]! \, g_{\mu_1 \cdots\mu_{d-p-1}}g^{\mu_1 \cdots\mu_{d-p-1}}
+ (-1)^s [d-p]!\,f_{\mu_1 \cdots\mu_p}f^{\mu_1 \cdots\mu_p}) .
\label{rho}
\ee

For a more concise notation, in terms of forms, consider the following definitions:
 $d(f , g) \equiv (df\, , dg)$, and
\be \mbox{}^{*}\, (df \, , ~ dg) \equiv (\mbox{}^{*}dg ~
,~(-1)^{p+1}\, S_{p+1} ~\mbox{}^{*}df)\,, \ee where $S_q$ is a
number defined by
 the double dualisation operation, for a $q$-form $A$:
$\mbox{}^{*}(\mbox{}^{*}A) = S_q \, A \,,$ this depends on the
signature ($s$) and dimension of the space time in the form $S_q
=(-1)^{s+q[d-q]}$.

Notice that $\mbox{}^{*}$ applied to doublets is defined such that
 its components are interchanged
 with a supplementary change of sign for the second component.

In so doing, the equations of motion derived from the action (\ref{DSD})
read as
\be
\label{sdrel}
{\cal F} = {1 \over  m }\,\mbox{}^{*} d{\cal F}   ,
\ee
where $m$ is a mass parameter introduced for dimensional
 reasons. It may trivially be verified that these equations require
 that ${\cal F}$ satisfies a Proca equation with mass $m$.

Notice that the equation (\ref{sdrel}) looks like (\ref{190}).
In that sense, we state that $S_{DSD}$
describes {\it doublet}-self-duality.

In the previous section, we considered non-linear generalizations of SD models;
 in the same sense,
we may replace $\rho$ by $V(\rho)$ in the action (\ref{DSD}) and obtain non-linear
 generalizations
of the model. Below, we are going to prove that these theories are dual equivalent to also
 non-linear
 topologically massive ones. The form of this correspondence shall result the same as to
 the $3-d$
case (Eq. (\ref{UV})), which constitutes an additional motivation
to interpretate (\ref{DSD})
 as a Self-Dual system.

Consider the doublet of gauge fields $ {\cal A} \equiv (a_{\mu_1 \cdots\mu_p} ,  b_{\mu_1 \cdots\mu_{d-p-1}} )$
in addition to $ {\cal F} =(f_{\mu_1 \cdots\mu_p} ,  g_{\mu_1 \cdots\mu_{d-p-1}} ) $;
we now propose the following parent action:
\bea
{\cal S}_{P}[{\cal A} , {\cal F}] = {\cal S}_{BF}[{\cal A} ] - \int dx^d \, \epsilon^{\mu_1\cdots \mu_{d}}
\left[ b_{\mu_1\cdots\mu_{d-p-1}} \partial_{\mu_{d-p}} f_{\mu_{d-p+1}\cdots\mu_{d}}
+ g_{\mu_1 \cdots \mu_{d-p-1}} \partial_{\mu_{d-p}} a_{\mu_{d-p+1}\cdots\mu_{d}}\right] + \nonumber\\
+ V(\rho({\cal F}))~],
\label{masterDJdob}
\eea
where
\be
\label{BF}
{\cal S}_{BF}[{\cal A}] \equiv \int dx^d \,\left[ b_{\mu_1\cdots\mu_{d-p-1}}
\epsilon^{\mu_1\cdots \mu_{d}} \partial_{\mu_{d-p}} a_{\mu_{d-p+1}\cdots\mu_{d}} \right]
\ee
is the BF-action.

Varying ${\cal S}_{P}$ with respect to ${\cal F}$, we obtain
 \be
{\cal F}= - {1 \over  V'(\rho)  }\,\mbox{}^{*} d{\cal A};
 \ee
which looks like non-linear self-duality, equation (\ref{AIS01B}).
Plugging this ralation back into (\ref{masterDJdob}),
we recover the generalized (non-linear) topologically massive action:
\be
\label{GMCSU}
 {\cal S}_{TM}[{\cal A}]= {\cal S}_{BF}[{\cal A}]  - \int d^{d}x \, \,U(\theta) ,
\ee where $\theta$ encodes the Maxwell-type terms: \be \theta
\equiv \frac{1}{2} \left( (-1)^s \,[d-p-1]!
\,(\partial_{[\mu}a_{\mu_1 \cdots\mu_p ]})^2 + \,[p+1]!\,
(\partial_{[\mu}b_{\mu_1 \cdots\mu_{d-p-1}]})^2  \right). \ee

Thus, the same algebraic manipulations that in 3d-case lead to
relate $U$ and $V$ again in terms of $V$ by Eq. (\ref{UV}),

We shall observe that this is invariant
 under the gauge transformations; $ {\cal A}  \to {\cal A}  +
 d{\cal  D}$, where $d{\cal D}$ is a {\it pure gauge doublet}, i.e, it is a pair of
exact differentials of $(p-1,d-p-2)$-forms.

\vspace{0.5cm}

Now, we vary ${\cal S}_{P}$ with respect to ${\cal A}$ and obtain:
 \be
\mbox{}^{*}d({\cal A}-{\cal F})=0 ;\ee
or in components,
\ba
\mbox{}^{*}d(a-f)&=&0\nonumber\\
\mbox{}^{*}d(b-g)&=&0\, .
\ea
 This implies that  the differences $a-f$ and $b-g$ may locally be
  written as exact forms; therefore,
 one it is possible to express the solution to these equations as
\be  {\cal A} = {\cal F} + d{\cal D} .\ee Putting this back into the
action (\ref{masterDJdob}) , we recover the generalized SD theory up to
topological terms:

\be \label{VDSD} S_{DSD}[{\cal F}]
\equiv \int dx^d \,\left[ - {1\over m} g_{\mu_1\cdots\mu_{d-p-1}}
\epsilon^{\mu_1\cdots \mu_{d}} \partial_{\mu_{d-p}}
f_{\mu_{d-p+1}\cdots\mu_{d}} + V(\rho({\cal F})) \right] .\ee

Whenever $V$ (or $U$) is linear, we get the so-called
 Cremmer-Scherk-Kalb-Ramond-type models, and the present result
reproduces the dual correspondence obtained by Harikumar et al in
the recent work of Ref. \cite{HS} for $d=3+1$, recently
generalized, in \cite{tmdob}, to arbitrary dimensions and all
possible tensorial ranks.

%%%%%%%%%%%%%%%%%%%%%%%%%%%%%%%%%%%%%%%%%%%%%%%%%%%%%%%%%%%%%%%%%%%%%%%%%%%%%%%

\subsection{More general non-linearities.}

It is not a general fact that $V=V(\rho({\cal F}))$. Besides the requirement
 of Lorentz invariance,
one may also require that
the two gauge forms which compose the doublet do not interact with one another,
apart from the interaction due to the BF-term.

Consider ${\cal F} \equiv (f_1 , f_2)$ and ${\cal A} \equiv (a_1,
a_2)$, both in $\Delta_p $, and the non-linearity given by
\be V=V_1(N_2\;(f_1)^{2} ) + (-1)^s V_2( N_1\;(f_2)^{2}) ;\ee
where $N_i \equiv \frac{[p_i +1]!}{2}$, $i=1,2$ and $p_i$ denotes the rank of $f_i$
($p_1+p_2 +1=d$) \footnote{ $(f_i)^2$ denotes
$f_{\mu_1\cdots\mu_{p_i}}f_{\mu_1\cdots\mu_{p_i}}$.}.

Then, the variation of (\ref{masterDJdob}) with respect to  ${\cal
F}$ yields: \be \left(\,V'_1(N_2\,(f_1)^{~2} )\; f_1\; ;\; V'_2(N_1\,(f_2)^{~2} )\;
 f_2 \,\right)= -\,\mbox{}^{*} d{\cal A} \, .
\ee
Thus, by repeating the previous calculations, we can
readily check the duality between

\be
 {\cal S}_{V_1 , V_2}[{\cal F}]={\cal S}_{BF}[{\cal F}]+
 \int\,d^{d}x\, \left(  V_1(N_2\,(f_1)^{~2} ) + V_2(N_1\, (f_2)^{~2} ) \right) ~~~~~~,
\label{GSDMdif} \ee
and
\be
\label{GMCSUdif}
 {\cal S}_{U_1 , U_2}[{\cal A}]={\cal S}_{BF}[{\cal A}]-
  \int\,d^{d}x\, \left( \; U_1( p_2!\, (d a_1 )^2 )
   + U_2( p_2!\, (d a_2 )^2 ) \; \right)\, ~~~~~~ .
\ee
Thus, we come to the known relation:
\be
\label{2UV}
U_i (q )=- 2 W_i (q ) V_i ' \,(W_i (q )) \, +   V_i (W_i (q ) )\,, ~~q  \ep R^+ ~;
\ee
 where, the functions $W_i $ are again defined by
\be W_i ^{-1}(v ) \equiv  2 v  [V_i '(v )]^2 ~~v  \ep R^+ ~.\ee

 \vspace{0.5cm}

In $d=3+1$, an interesting duality can be established by applying this result
to a topologically massive combination of a Born-Infeld with a
(rank two) Kalb-Ramond field.

The Born-Infeld-Kalb-Ramond theory \footnote{Here, Born-Infeld
means that the free action is proportional to $\sqrt{1- const.
F_{\mu \nu}F^{\mu \nu}}$.} \be \label{par2} S_{BIKR}({\cal A})
=\int d^{4}x\left( \beta^2\sqrt{[1- \frac{2}{\beta^2} \partial_{[
\rho} A_{\mu ]}\partial^{[ \rho} A^{\mu ]}] }- \partial_{[ \rho}
B_{\mu \nu ]}\partial^{[ \rho} B^{\mu \nu]} + m
B_{\mu\nu}\epsilon^{\rho\mu\nu\sigma}\partial_{\rho}A_{\sigma}\right),
\ee for the doublet of gauge fields ${\cal A}= (A_\mu \, , B_{\mu
\nu})$ \footnote{$\beta$ is a parameter introduced for dimensional
reasons.} , is dual-equivalent to the first-order model: \be
\label{dob2k} S_{DSD}({\tilde{\cal A}}\equiv({\tilde
A}_{\sigma},{\tilde B}_{\mu\nu}))=
 \int d^{4}x \left(- \beta^2\sqrt{[1+ \, \frac{1}{\beta^2}
{\tilde A}_{\sigma}{\tilde A}^{\sigma}]} +  {\tilde B}_{\mu\nu}{\tilde B}^{\mu\nu}
+{1\over m} {\tilde A}_{\sigma}\epsilon^{\sigma\rho\mu\nu}
\partial_{[ \rho}{\tilde B}_{\mu\nu ]}  \right) ,
\ee which is a gauge non-invariant theory, also associated to a non-linear
doublet-self-duality constraint.

\subsection{Bosonization in (3+1)-d}

Here, we present a novel approach to bosonization in $d=3+1$, valid
for length scales long compared with the Compton wavelength of the fermion.

In a four dimensional massive fermionic model with $U(1)$ charge;
just like in the 3d-case, one defines a current: $j^\mu \equiv
\bar{ \psi} \gamma^\mu
 \psi$, where $\psi$ are $N_f$ four-component Dirac spinors
 \footnote{In this calculation, $N_f$ will be simply considered as a
 parameter.}.

However, one can also define a rank-two current, $j^{\mu \nu}
\equiv  \, \bar{ \psi} \gamma_5 [\gamma^\mu \, ,\gamma^\nu
\,] \psi $;
 let us now define the doublet-current:
\be {\cal J} =(j^\mu , j^{\mu \nu} ). \ee The appearance of the
$\gamma_5$-matrix in $j^{\mu \nu} $ follows from requiring
that $j^{\mu \nu}$ as well as $j^{\mu } $ are both odd
 under charge conjugation: $ \bar{ \psi} \gamma_5
[\gamma^\mu \, ,\gamma^\nu \,] \psi = -\bar{ \psi^C} \gamma_5
[\gamma^\mu \, ,\gamma^\nu \,] \psi^C $. Notice that ${\cal J}$
is a well-formed doublet $( {\cal J} \,\epsilon \, \Delta_p)$.

 Now, we can write a non-conventional (Euclidean)\footnote{In an Euclidean space-time,
 $j^{\mu \nu}$ is purely imaginary; thus, in order to render it real, we
 may redefine this bilinear by multiplying it by an $i$.} massive Thirring model in a
similar fashion to the 3d-case:

\be \label{ferdob} {\cal Z}^{(ferm)}\equiv
  \int \,{\cal D}\bar\psi {\cal D}\psi\; e^{-\int \left ( \bar\psi(\dslash
+ m) \psi - \frac{g^2}{2 N_f m}[2\,j_{\mu \nu}j^{\mu \nu} -
j_{\mu}j^{\mu}]\right) d^4 x} ,\ee
where $m$ is the fermion mass and $g$ a coupling constant of the model,
such that $g^2$ have dimensions of inverse mas.

We are going to show that this bosonizes into the CSKR-model, a
gauge, topologically massive theory.

 Such as in the 3d-case, we get the
identity, \be
e^{-\frac{g^2}{2\, N_f\,m} \int\!d^4\!x\, [2\,j_{\mu \nu}j^{\mu
\nu}- j_{\mu}j^{\mu}]}  = \int{\cal D} {\cal A}\;
e^{\int\!d^4\!x\, tr(\frac{1}{2} [\frac 12 b_{\mu \nu}b^{\mu \nu}-
a_{\mu}a^{\mu}] + \frac{g}{\sqrt{m N_f}} [ b_{\mu \nu}j^{\mu \nu}- a_{\mu}j^{\mu}])}
,\label{NB20dob} \ee
 which introduces the doublet of bosonic
fields ${\cal A} \equiv (a_{\mu } , b_{\mu \nu})$.

Defining the doublet-slash by \be \Aslash \equiv \gamma^\mu a_\mu
+ \gamma_5 [\gamma^\mu \, ,\gamma^\nu \,] b_{\mu \nu}\,\, , \ee the
partition function reduces to:

\be {\cal Z}^{(ferm)} = \int {\cal{D}} {\cal A} \; \det (i \dslash
+ m + \Aslash) \, e^{\frac 12 \!\int d^4 x \; [\frac 12 b_{\mu
\nu} b^{\mu \nu} - a_{\mu}a^{\mu}]} . \label{NB300} \ee

 Next, we must evaluate this determinant.

A straightforward perturbative expansion yields \be
S_{eff}[\Aslash,m]= N_f \, tr [\ln (\dslash + m)] +\, N_f\,\frac{g}{\sqrt{N_f\,m}}
tr\left({1\over  \dslash + m} \Aslash \right )+ \frac{N_f}{2}\,\left(\frac{g^2}{N_f\,m}\right)
tr\left({1\over  \dslash + m} \Aslash {1 \over  \dslash+m}
\Aslash \right)+\dots \ee

The first term is just the free $(A=0)$ case, which is subtracted,
while the second term are simply two tadpoles accommodated in the
doublet. Thus, we draw our attention to the quadratic term (in the
bosonic fields ${\cal A}$) in the effective action. In momentum
space, this reads

\be
 S_{eff}^{quad}[A,m]= \frac{g^2}{2 m}\, tr\, \int {d^4 p \over (2\pi)^4 }\,{d^4k\over (2\pi)^4 } \left[
\Aslash(-p) {i\pslash + i\kslash-m\over (p+k)^2+m^2}\,
\Aslash(p){i \kslash-m\over k^2+m^2} \right].
 \ee
 Terms of the form $\Aslash(-p)\kslash\Aslash(p)\kslash$
 and $\Aslash(-p)\pslash\Aslash(p)\kslash$ in the numerator
 of the integrand will contribute
 at most to second order in  $p_\mu $
 \footnote{They also cancel the terms $tr \, [ m^2 \Aslash(-p)\Aslash(p)]$
that appear in the numerator.}.
Since we are seeking for the low energy limit,
 like in the 3d-case, we can approximate this by

\be S_{eff}^{quad}[A,m] \approx  i\,\frac{g^2}{2 m}\,  tr \, \int
 {d^4p \over (2\pi)^4} \,{d^4k\over (2\pi)^4} \left[ \Aslash(-p)
 { (\pslash + \kslash) \Aslash(p)\, - \, \Aslash(p) \kslash
  \over [(p+k)^2 + m^2]\,  [k^2+m^2]} \right],
 \ee
using also the fact that the trace of an odd number of
 gamma-matrices is zero. We obtain only a topological contribution:

 \be
 S_{eff}^{quad}[A,m] \approx  \frac{g^2}{2}\,  \int {d^4p \over (2\pi)^4 }\, \left[
a_\mu(-p) \Gamma^{\mu\nu\alpha}(p) b_{\nu \alpha}(p)\right],
\label{quadratic}
\end{equation}
where, by virtue of the special property of the gamma
  matrices (here, Euclidean) in $(3+1)$-d ,
\be tr(\gamma^\mu \gamma^\nu \, \gamma_5 [\gamma^\rho \,
,\gamma^\alpha \,])= - 8\epsilon^{\mu\nu\rho\alpha}, \label{trodd}
\ee
 the kernel takes the form:
\begin{equation}
\Gamma^{\mu\nu\alpha}(p,m)=  \epsilon^{\mu\rho\nu\alpha}p_\rho
\Pi(p^2,m), \label{form}
\end{equation}
where $\Pi(p^2,m)$ is the contribution corresponding to
 the one-fermion-loop self-energy diagram.
For the sake of computing the loop integral and factoring out
the divergent part, we go over to $d=4-\ep$-dimensions,
following the procedure of dimensional regularization
(see ref. \cite{hel}):

\begin{eqnarray}
\Pi(p^2,m)&=& (\mu)^{\ep}\int{d^d k\over (2\pi)^d}
{1\over [(p+k)^2+m^2][k^2+m^2]}\nonumber\\
 &=&\frac{1}{(4\pi)^2}\left[\frac{2}{\ep}-\gamma -ln \frac{p^2}{\mu^2}
-I(\frac{p^2}{m^2} )\right] + o(\ep)\, ,
\label{arcs}
\end{eqnarray}
$\mu$ is a parameter and the finite part reads as below: \be
I\left(\frac{p^2}{m^2} \right)= a \, ln a - (a -1) \, ln (a -1)
 + b \, ln |b| + (1- b )\, ln (1- b ) - 2 ,
\ee where
\bea
a &=& \frac{1}{2} \left[1 \pm \sqrt{1+ 4\frac{m^2}{p^2}} \right]\;,\nonumber\\
b &=& \frac{1}{2} \left[1 \pm \sqrt{1- 4\frac{m^2}{p^2}} \right].
\eea In the long wavelength ($p\to 0$) and large mass
($m\to\infty$) limit, $a \to \infty\, ,\, b \to -\infty $; thus,
it is easily verifiable that $I\to -2$. Therefore, we find the
finite part of the kernel:

\begin{equation}
\Gamma^{\mu\nu\alpha}(p,m)\sim \,\frac{2}{(4\pi)^2}\,
\epsilon^{\mu\rho\nu\alpha}p_\rho
 \label{exp}
\end{equation}

Inserting the leading term into the quadratic effective action
(\ref{quadratic}) and going back to configuration space
(Lorentzian), we find an induced BF-term

\begin{equation}
\label{bfind}
S_{eff}= - 8 \frac{g^2}{(4\pi)^2}\, \int d^4 x
\epsilon^{\mu\nu\rho\alpha}a_\mu \partial_\nu b_{\rho\alpha}=
- 8 \frac{g^2}{(4\pi)^2}\, {\cal S}_{BF} ({\cal A}). \label{indquad}
\end{equation}
Putting this result back into (\ref{NB30}), we obtain:

\be
{\cal Z}^{(ferm)}\approx \int {\cal{D}} {\cal A} \; e^{{\cal S}_{BF}  ({\cal
A})+ \frac{1}{2}\!\int d^4 x\: [\frac 12 b_{\mu \nu}b^{\mu \nu}-
a_{\mu}a^{\mu}]} ,
 \ee
which, via the correspondence proven before, is
equivalent to the gauge invariant Cremer-Sherk-Kalb-Ramond model
which describes a massive spin-one (bosonic) particle.
 The boson mass is given by the inverse of the factor in front of ${\cal S}_{BF}$
in the eq. (\ref{bfind}), $m_{boson}^{-1} \sim   \frac{g^2}{2 \pi^2}$.

Notice also that, if one rescale the doublet current as
$(j^\mu , j^{\mu \nu} )\to(s j^\mu , t j^{\mu \nu} )$,
the single effect of this that the boson mass
results rescaled as, $m_{boson} \to m_{boson}\, /\, (st)$.

 Finally, a {\it fermionic representation} for the
CSKR model is given by the partition function: \be {\cal
Z}^{(ferm)} =
  \int \,{\cal D}\bar\psi {\cal D}\psi\; e^{-\int \left ( \bar\psi(\dslash
+ m) \psi -\frac{g^2}{2\, N_f\,m}  \left[2\, j_{\mu \nu}j^{\mu \nu}+
j_{\mu}j^{\mu}\right] \right) d^3x} \approx {\cal Z}_{CSKR} .\ee

Now, by repeating the calculations of the previaous subsections, one
may to study non-linear generalizations of the fermionic model
(\ref{ferdob}). In fact, substituting $j_{\mu \nu}j^{\mu \nu}+
j_{\mu}j^{\mu} \;\to\;U_1(j_{\mu \nu}j^{\mu \nu})+U_2(j_{\mu}
j^{\mu})$ in the expression (\ref{ferdob}), one can bosonize
this into a non-linear SD theory given by (\ref{GSDMdif}) \footnote{For
simplicity, we are discussing on the case $d=4$ and doublets in
$\Delta_1$.}, whose non-linearities are related to $U_{1/2}$ by the
expressions (\ref{2UV}). And once more, for composing this with the duality
proven in the subsection 4.1 , this corresponds to a topologically
massive gauge theory (so as in the Thirring-MCS correspondence)
given by the action (\ref{GMCSUdif}).

In particular, we can write down the fermionic counterpart of the
Born-Infeld-Kalb-Ramond  gauge theory. This may be cast as
 \be
{\cal Z}_{BI-KR} \approx \int \,{\cal D}\bar\psi {\cal D}\psi\;
e^{-\int \left ( \bar\psi(\dslash + m) \psi - \frac{g^2}{2\, N_f\,m}
[2 j_{\mu\nu}j^{\mu\nu} + \beta^2 \sqrt{1-\frac{j_{\mu}j^{\mu}}{\beta^2})}]\right )
d^3x} .\ee

Let us conclude this section by mentioning that the operator
correspondence underlying this structure reads as
\be
{\cal J}~\to~\mbox{}^{*} d{\cal A}\, .
\ee

%%%%%%%%%%%%%%%%%%%%%%%%%%%%%%%%%%%%%%%%%%%%%%%%%%%%%%%%%%%%%%%%%%%%%%%%

\section{Final Remarks.}

We have presented here a new approach to study the bosonization of a
model of interacting fermions in terms of topologically massive models,
similar to what happens in $d=3$. In general,
this involves two gauge fields with different tensorial
 ranks (BF-type theories). We have actually discussed this point for $d=4$,
but we showed the road to reproduce this construction in higher
dimensions (one simply should build up the currents as elements
 in some $\Delta_p$).
 These results have been emphasized for theories which
appear to be very important in field theory and/or dynamics of
$Dp$-branes (CSKR and BIKR theories).

 A comment is in order that regrads the two-form current, $j^{\mu \nu}$, appearing in the Thirring
 model. It may look somewhat artificial, since it is  not
  necessarily conserved.
 Nevertheless, we try here to show that it is actually
  a natural piece of the formalism,
  since it is
 related to topologically massive gauge invariant models:
 it is crucial for the attainment of a bosonic topologically massive
 theory in the large fermionic mass limit. Bosonization in the case of
 non-conserved fermionic currents has already been contemplated
 by other authors \cite{cortes}.

We conclude this paper by stressing a motivation for the proposed
generalization of the self-duality to $d>3$, via doublets
\cite{dob,tmdob}. It appears to be appropriate to highlight such a point,
since, despite the use of the doublet procedure
 proposed here to bosonize a $3d$-Thirring model,
 one recovers the well-known results in $3d$,
 i.e; the doublet disappears and reduces to a single dynamic
self-dual field. In fact, for a Thirring model in $3d$, with a
$U(1)$-interaction, we can only construct a current doublet in
$\Delta_1$,  ${\cal J}=(j^\mu ,  j^{\mu} )$, where $j^{\mu}
\equiv \bar{ \psi} \gamma^\mu \,  \psi $. After introducing, as
usual, a bosonic doublet ${\cal A}=(a_\mu , b_{\mu } )$,
 the partition function may be cast as
\be {\cal Z}_{(ferm)} \equiv
  \int \,{\cal D}\bar\psi {\cal D}\psi\, {\cal D}a {\cal D}b\;
 e^{-\int \left( \bar\psi(\dslash
+ m) \psi -\frac{g^2}{2}j^\mu\,[a_\mu + b_\mu] - (a^2 + b^2)/2
\right ) d^4 x} \, .\ee By changing coordinates to $c_\mu^{\pm}
\equiv {a_\mu \pm b_\mu \over 2} $, the field $c_\mu^+$ appears
decoupled from $c_\mu^-$ (the latter without dynamics), whose
 action, induced by the fermionic model,
is precisely given by a self-dual model (eq. (\ref{180})), as
expected. This fact seems to be an additional motivation to think of
the (current) doublets as more general objects.

{\bf Aknowledgements}: The authors express their gratitude
to CNPq, for the invaluable financial help.

\end{document}